\documentclass[11pt]{article}
\usepackage{moriond,epsfig}

\def\Journal#1#2#3#4{{#1} {\bf #2}, #3 (#4)}

\def\PLB{{\em Phys. Lett.}  B}

\def\PRD{{\em Phys. Rev.} D}

\def\beqra{\begin{eqnarray}}
\def\eeqra{\end{eqnarray}}
\def\beq{\begin{equation}}
\def\eeq{\end{equation}}

\def\vp{\varphi}

\begin{document}
\vspace*{4cm}
\title{Enhancement of Dark Matter Abundance in Scalar-Tensor Dark Energy}

\author{R.~CATENA$^{(1)}$, N.~FORNENGO$^{(2)}$, A.~MASIERO$^{(3,4)}$, M.~PIETRONI$^{(4)}$ and F.~ROSATI$^{(3,4)}$  }

\address{$^{(1)}${\it Scuola Normale Superiore and INFN - Sezione di Pisa, \\
Piazza dei Cavalieri 7, I-56125 Pisa, Italy }\\
$^{(2)}${\it Dipartimento di Fisica Teorica, Universit\`a di Torino and \\ INFN - Sezione di Torino, 
via P. Giuria 1 I-10125 Torino, Italy  }  \\ $^{(3)}${\it Dipartimento di Fisica, Universit\`a di Padova,  via Marzolo 8, I-35131, Padova, Italy } \\  $^{(4)}${\it INFN, Sezione di Padova, via Marzolo 8, I-35131, Padova, Italy} \\ }

\maketitle\abstracts{
In Scalar-Tensor theories of gravity, the expansion rate of the universe in the past may differ considerably from the standard one.  After
imposing the constraints coming from nucleosynthesis, CMB, type Ia supernovae, and solar system tests of General Relativity, we consider the expansion rate of the universe at WIMP decoupling, showing that it
can lead to an enhancement of the dark matter relic density of some
orders of magnitude with respect to the standard case. This effect may have a deep impact on most popular candidates for dark matter, as the supersymmetric neutralino.
}

%
Dynamical models of Dark Energy (DE), in which the energy component responsible for the acceleration of the universe evolves on a cosmological time-scale, generically require ultra-light degrees of freedom, with masses of order $H_0\sim 10^{-33}$ eV.
Scalar-tensor (ST) theories represent a natural framework in which massless scalars
may appear in the gravitational sector of the theory without being
phenomenologically dangerous. In these theories a metric coupling of
matter with the scalar field is assumed, which preserves the Einstein equivalence principle.

Moreover, as discussed in \cite{dam3}, a large
class of these models exhibit an attractor mechanism towards General Relativity (GR), that
is, the expansion of the Universe during the matter dominated era
tends to drive the scalar fields toward a state where the theory
becomes indistinguishable from GR.

ST theories of gravity may be defined by the `Einstein frame'  gravitational action 
\begin{equation}
S_{g}=\frac{M_{\ast}^2}{2}\int d^{4}x\sqrt{-{g}}\left[ {R}+{g}^{\mu
\nu }\partial _{\mu }\varphi \partial _{\nu }\varphi -\frac{2}{M_{\ast}^2} V(\varphi )\right] ,
\end{equation}
and the matter action, where the scalar field $\vp$ appears  through a purely metric coupling, 
\begin{equation} 
S_m = S_{m}[\Psi_{m},A^{2}(\varphi ){g}_{\mu \nu }] \,\,\, . 
\end{equation}
Notice that, in this frame, masses and non-gravitational coupling constants are space-time dependent, and that the energy-momentum tensor of matter fields is
not conserved separately, but only when summed with that of the scalar field. On the other hand, an observer measuring time and lengths with non-gravitational ({\it e.g.} atomic) clocks and rods would be coupled to the `physical' metric
\beq
\tilde{g}_{\mu\nu} = A^{2}(\varphi )g_{\mu \nu }\,,
\eeq
and will not measure any space-time dependence of masses and non-gravitational couplings, while she/he  will generally see some differences in the gravitational sector. We chose to solve the equations of motion in the Einstein frame, because they are simpler, and we then translate the results in the physical frame to compare them with observations, or with the non-gravitational rates responsible for the WIMP annihilation and nucleosynthesis. 

If $A(\vp)$ were a constant, the action above would be  undistinguishable  from the standard one. A crucial quantity, measuring the `distance' from GR,  is then
\beq
\alpha (\varphi ) \equiv \frac{d\log A(\varphi )}{d\varphi }\,.
\eeq
$\alpha \to 0$ being the GR limit.
The Friedmann equations in the Einstein frame have the usual form \cite{noi} while the scalar field evolves according to the equation
\beq
\ddot{\varphi} +3\frac{\dot{a}}{a}\dot\varphi =
-\frac{1}{M_{\ast}^2}\left[ \frac{\alpha (\varphi )}{\sqrt{2}}(\rho-3p)+  \frac{\partial V}{\partial \varphi } 
 \right]
 \label{eom}
 \eeq
One of the relevant quantities in the discussion of the WIMP freeze-out is the physical expansion rate, which is given by
\begin{equation}
\tilde{H}^2 = \frac{A^2(\vp )}{3M_*^2} \, 
\frac{\left( 1 + \alpha (\vp )\,\vp ^{\prime} \right) ^2}{1-(\vp ^{\prime})^2 /6} \, 
\left[\tilde {\rho} + \tilde {V} \right] \,\,\,\, ,
\label{Htilde2}
\end{equation} 
where primes denote differentiation with respect to the logarithm of the Einstein frame scale factor.
 

%
%
%
At first sight, during radiation domination the scalar field would not move, since the potential energy density is
suppressed while the first term in the RHS of
Eq.~(\ref{eom}) vanishes due to the relativistic equation of state, $\rho=p/3$. However,  when the temperature is around a particle's mass, the combination $\rho-3 p$ for that particle is momentarily sizable (before becoming Boltzmann suppressed for $T<m$), and gives a `kick' to the field evolution. This effect is crucial for what follows, since it allows an effective evolution of the system from a situation far away from GR at the WIMP freeze-out temperature, to one much closer to it at nucleosynthesis.

During matter domination $\rho-3p \simeq \rho$ and, as long as the potential
energy density is subdominant  the field evolution depends on the form of the coupling function
$\alpha(\varphi)$.  

It was emphasized in Ref. \cite{max} (see also \cite{sabino2}) that the fixed point starts to be
effective around matter-radiation equivalence, and that it governs the
field evolution until recent epochs, when the quintessence potential
becomes dominant. If the latter has a run-away behavior, the same
should be true for $\alpha(\varphi)$, so that the late-time behavior
converges to GR.

The evolution of the field during the last redshifts depends on the
nature of DE. We considered two possibilities: a cosmological
constant and a inverse-power law scalar potential for $\varphi$.



Now we come to the discussion on the phenomenological bounds on this model. From the expression of the expansion rate in Eq.~(\ref{Htilde2}) we see that a variation of $A^2(\vp)$ can be interpreted as a variation of the Newton constant. Imposing that the maximum allowed variation of  expansion rate during Nucleosynthesis  with respect to the standard case is that equivalent to an extra neutrino family, we get the bound
\beq
\frac{A(\vp_{nuc})}{A(\vp_0)} <1.08\,.
\label{BBN}
\eeq
Then we consider the post-newtonian limits coming from GR tests in the solar system. Using the new results on time delays of radio signals obtained from the Cassini spacecraft we obtain the following bound on the value of the function
$\alpha$  today,
\beq
\alpha^2(\vp_0)=(-1.0\pm 1.2) \times 10^{-5}.
\label{casbound}
\eeq
The bound from the Cassini spacecraft turns out to be quite strong
when used in connection with the one on the equation of state $w_\vp$
from SNe Ia~\cite{SNe} ( $w_\vp < -0.78$ at 95\% c.l.). Indeed, the more the equation of state is close to the cosmological constant value
 $w_\vp = -1$,  the shallower the scalar potential must be. This implies a very small evolution  of the function  $\alpha$ during the last redshifts, so that for $w_\vp \to 1$ it would be increasingly difficult to see any deviation from GR from observations of the cosmic expansion in the recent past.
 
A variation of  the gravitational constant at radiation decoupling would in principle have an impact on the locations of the peaks of the CMB power spectrum. However we have checked that, once the bounds on BBN and GR have been imposed, the field evolution after BBN  is strong enough as to bring
the scalar tensor theory at recombination so close to GR as to leave no signatures on the CMB 
spectrum.

Let me finally show  the results of the calculation of the relic abundance of a DM WIMP
with mass $m$ and annihilation cross-section $\langle\sigma_{\rm ann}
v\rangle$. As already mentioned, laboratory clocks and rods measure
the ``physical'' metric $\tilde{g}_{\mu\nu}$ and so the standard laws
of non-gravitational physics take their usual form in units of the
interval $d\tilde{s}^2$. As outlined in Ref.\cite{damour-pichon}, the
effect of the modified ST gravity will enter the computation of
particle physics processes (like the WIMP relic abundance) through the
``physical'' expansion rate $\tilde{H}$ defined in Eq.~(\ref{Htilde2}).
We have therefore to implement the standard Boltzmann equation with
the modified physical Hubble parameter $\tilde{H}$:
\begin{equation}
\frac{dY}{dx} = -\frac{1}{x} \frac{s}{\tilde{H}} \langle\sigma_{\rm
ann} v\rangle (Y^2 - Y_{\rm eq}^2)
\label{eq:boltzmann}
\end{equation}
where $x=m/T$, $s=(2\pi^2/45)~h_\star(T)~T^3$ is the entropy density
and $Y=n/s$ is the WIMP density per comoving volume.

The solution of the Boltzmann equation is formally the same
as in the standard case, with the noticeable difference that now the
Hubble parameter gets an additional temperature dependence, given by
the function $A(\varphi)$. This can be translated in a change in the
effective number of degrees of freedom at temperature $T$.
\begin{figure*}[t] 
\epsfxsize=3. in 
\epsfbox{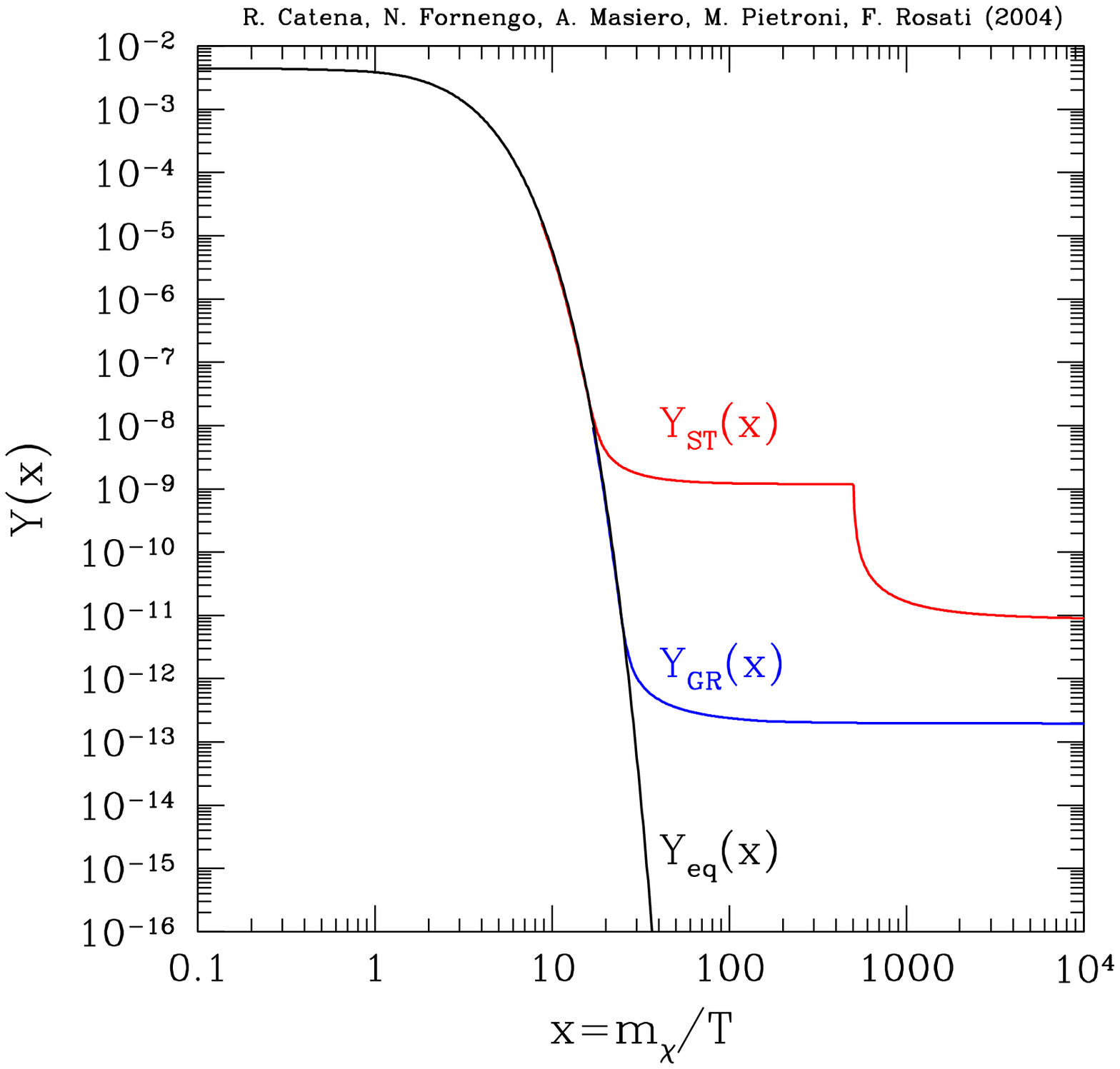}
\epsfxsize=3. in 
\epsfbox{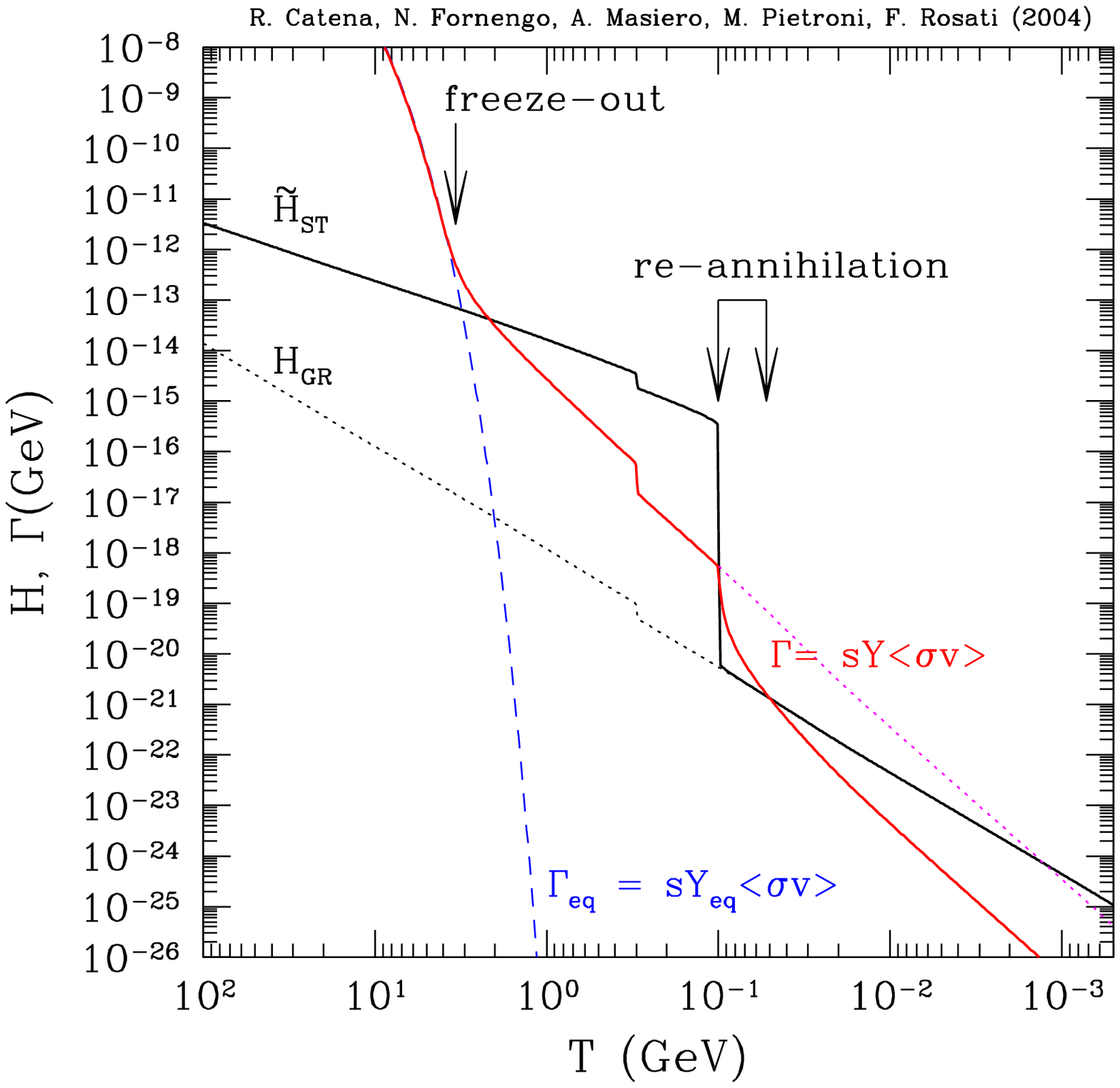}
\caption{\label{fig:abundance}(left) Numerical solution of the Boltzmann
equation Eq.~(\ref{eq:boltzmann}) in a ST cosmology for a toy--model
of a DM WIMP of mass $m=50$ GeV and constant annihilation
cross-section $\langle\sigma_{\rm ann} v\rangle = 1\times 10^{-7}$
GeV$^{-2}$. The temperature evolution of the WIMP abundance $Y(x)$
clearly shows that freeze--out is anticipated, since the expansion
rate of the Universe is largely enchaced by the presence of the scalar
field $\varphi$ (right). At a value $x=m/T_\varphi$ a re--annihilation phase
occurs and $Y(x)$ drops to the present day value.}
\end{figure*}
A numerical solution of the Boltzmann equation Eq.~(\ref{eq:boltzmann})
in a ST cosmology is shown in Fig.~\ref{fig:abundance} for a
toy--model of a DM WIMP of mass $m=50$ GeV and constant annihilation
cross-section $\langle\sigma_{\rm ann} v\rangle = 1\times 10^{-7}$
GeV$^{-2}$. The temperature evolution of the WIMP abundance $Y(x)$
clearly shows that freeze--out is anticipated, since the expansion
rate of the Universe is largely enchaced by the presence of the scalar
field $\varphi$. This effect is expected. However, we note that a
peculiar effect emerges: when the ST theory approached GR (a fact
which is parametrized by $A(\vp) \rightarrow 1$ at a temperature
$T_\varphi$, which in our model is 0.1 GeV), $\tilde H$ rapidly drops
below the interaction rate $\Gamma$ establishing a short period during
which the already frozen WIMPs are still abundant enough to start a
sizeable re--annihilation. This post-freeze--out ``re--annihilation
phase'' has the effect of reducing the WIMP abundance, which
nevertheless remains much larger than in the standard case. For the
specific case shown in Fig.~\ref{fig:abundance} the WIMP relic
abundance is $\Omega h^2=0.0027$ for GR, while for a ST cosmology
becomes $\Omega h^2=0.12$, with an increase of a factor of 44.


Needless to say, such potentially  
(very) large deviations entail new prospects on the WIMP characterization
both for the choice of the CDM candidates and  for their direct and indirect
detection probes. A thorough reconsideration of the ``traditional''
WIMP identified with the lightest neutralino in SUSY extensions of the SM
as well as the identification of other potentially viable CDM candidates
in the ST context  is presently under way. 

Similar effects on the relic abundances as those presented in this talk can be induced by a phase of kinetic energy dominance during the WIMP freeze-out \cite{Profumo-Ullio}.

\end{document}